\documentclass[12pt]{article}

\usepackage{lineno}
\usepackage[export]{adjustbox}
\usepackage[margin=1in]{geometry}
\usepackage{amssymb,amsmath,latexsym}                        
\usepackage{graphicx}
\usepackage{cite}
\usepackage{xcolor}
\usepackage{float}

\usepackage{amsthm}
\usepackage{multicol}
\usepackage{authblk}
\usepackage{subfigure}

\usepackage[hyperfootnotes=false]{hyperref}
\hypersetup{
 colorlinks=true,
 citecolor=blue,
 linkcolor=blue,
 urlcolor=blue}

\usepackage{adjustbox}
\usepackage{bbm}
\usepackage{mathtools}

\usepackage[titletoc]{appendix}
\date{\today} 

\begin{document}

\title{Nonlocal Thresholds for Improving the \\ Spatial Resolution of Pixel Detectors}
\author[1,2]{Benjamin Nachman}
\author[3]{Alex Spies}
\affil[1]{\normalsize\it Physics Division, Lawrence Berkeley National Laboratory, Berkeley, CA 94704, USA}
\affil[2]{\normalsize\it Simons Institute for the Theory of Computing, University of California, Berkeley, Berkeley, CA 94720, USA}
\affil[3]{\normalsize\it Department of Physics, University of California Berkeley, Berkeley, CA 94720, USA}

\maketitle

\begin{abstract}
Pixel detectors only record signals above a tuned threshold in order to suppress noise.  As sensors become thinner, pitches decrease, and radiation damage reduces the collected charge, it is increasingly desirable to lower thresholds.  By making the simple, but powerful observation that hit pixels tend to be spatially close to each other, we introduce a scheme for dynamic thresholds.   This dynamic scheme can enhance the signal efficiency without significantly increasing the occupancy.  In addition to presenting a selection of empirical results, we also discuss some potential methods for implementing dynamic thresholds in a realistic readout chip for the Large Hadron Collider or other future colliders.
\end{abstract}

%-------------------------------------------------------------------------------
\section{Introduction}
\label{sec:intro}
%-------------------------------------------------------------------------------

Pixel detectors are designed to be thin, to be highly granular, and to have low occupancy in order to precisely reconstruct charged-particle trajectories (tracks) from minimum ionizing particles (MIPs).   In order to achieve this goal while maintaining a high signal efficiency, only signals above a tuned threshold are recorded.  This threshold is chosen to be small compared to a typical signal, but large compared to the noise.  For example, sensors in the current Large Hadron Collider (LHC) experiments ATLAS and CMS are 200-300 $\mu$m thick, leading to a signal at perpendicular incidence of 16k-24k electrons (e); noise levels (measured as the equivalent noise charge or ENC) are typically 100-150e and tuned thresholds are  2k-3ke.  With these settings, the noise occupancy is well below $10^{-6}$~\cite{Abbott:2018ikt,Chatrchyan:2009aa}.

Given the increased instantaneous luminosity at the High Luminosity LHC (HL-LHC) and the goal of improving track reconstruction, there is a move towards thinner and narrower sensors.  Such sensors will require lower thresholds to compensate for the reduced signal charge resulting from the decreased path length of MIPs.  At the same time, the high particle flux expected at the HL-LHC also poses challenges due to increased radiation damage;  Non-ionizing energy loss results in defects in the sensor bulk that act as trapping sites and reduce the collected charge.  The large fluence also increases the noise via the sensor leakage current, $\text{ENC}_\text{leak}\propto \sqrt{I_\text{leak}}$ and $I_\text{leak}\propto\Phi$ (aside from annealing effects)~\cite{Rossi:976471}.  Indeed, with the current LHC lifetime fluence at circa $10^{15}$ 1 MeV $n_\text{eq}/\text{cm}^{2}$~\cite{Nachman:2649731}, the charge collection efficiency has dropped to roughly 70\%~\cite{Bomben:2650494} and the leakage current has reached one mA or more~\cite{PIX-2016-006,PIX-2018-008,cmsleakage}.  The ATLAS and CMS collaborations are working together within the RD53 collaboration~\cite{Garcia-Sciveres:2113263} to develop a new readout chip for their HL-LHC pixel detectors and therefore now\footnote{Given that both ATLAS and CMS are designing their innermost layers to be replaceable, new ideas may still see utilization in subsequent years, even if they are not fully developed in time for the upcoming production runs.} is a critical time to find solutions that address, at least in part, the challenges associated with the next-generation of pixel designs.
%see also p22.
%$\text{ENC}_\text{leak}\propto \sqrt{I_\text{leak}}$ and $I_\text{leak}\propto\Phi$. %p175
To this end, we propose a new method for pixel thresholding which stems from a simple, but significiant observation about MIP and noise hits: while the probability for a single pixel to be hit by a MIP is 0.1\% or smaller~\cite{Collaboration:2285585,Collaboration:2272264}, the probability for a pixel to be hit given that one of its neighboring pixels was hit is 10\% or more~\cite{ITK-2016-003}.  While the neighboring pixel hits can be caused by charge sharing from diffusion and capacitive coupling, they can also be due to an inclined primary particle traversing multiple sensors at an angle. Coupled with the fact that noise hits exhibit no spatial correlation,  this suggests that the optimal threshold should depend on the pattern of neighboring hits.  We study multiple implementations of this idea.

We are not aware of any previous efforts to utilize neighboring pixel information to dynamically adjust thresholds.  There have been previous proposals to implement dynamic thresholds to correct for spatial-temporal effects using information from a particular pixel~\cite{Garcia-Sciveres:2017pvu}.  A related topic is dual thresholds, which have been used extensively to separate time and energy measurements in order to make the best of both for a single detection.  This technique has been applied to precision timing ($\mathcal{O}(10\text{ ps})$) applications as diverse as positron emission tomography detectors~\cite{Rolo_2013} and high energy physics timing detectors~\cite{Collaboration:2296612,Albrow:1753795,MERLE201496,Rolo_2017} as well as `standard' LHC pixel detectors with timewalk concerns at $\mathcal{O}(10\text{ ns})$ timescales~\cite{790680}.  Dual thresholds have also been used for improving the position resolution by using one threshold for event triggering and one for measuring charge in regions of interest~\cite{5075458}.  Such a scheme is not possible for the extreme event rate at LHC pixel detectors, but the idea is similar to our proposal.  Another body of related work is dynamic and dual thresholds for edge detection (see e.g.~\cite{doi:10.1080/2150704X.2014.912766,Nain_dynamicthresholding,4310076}).  One of our proposals for implementing non-local thresholds will involve modifying the capacitative coupling between neighboring pixels.  This form of charge sharing has been well-suited in the literature (see e.g. Ref.~\cite{pixelbook}), but is traditionally viewed as a nuisance.  We show that this effect may instead be an asset for improving position resolution and increasing signal efficiency.  In contrast to traditional dual-threshold methods, this algorithm requires a fast communication between neighboring pixels and therefore has more stringent timing and power constraints.

This paper is organized as follows: Section~\ref{sec:simulation} introduces the simulation setup and Sections~\ref{sec:calculation} and~\ref{sec:schemes} introduce the metrics and threshold schemes, respectively;  The results are presented in Section~\ref{sec:results} with a brief discussion on implementation in Sec.~\ref{sec:disc}; the paper ends with conclusions and outlook in Section~\ref{sec:concl}.
 
%-------------------------------------------------------------------------------
\section{Simulation}
\label{sec:simulation}
%-------------------------------------------------------------------------------

A standalone simulation setup using Allpix~\cite{benoit:20xx} built on the Geant4 package~\cite{Agostinelli:2002hh} is used to simulate single particles interacting with a single planar pixel layer.  The sensor specifications are similar to those proposed for the ATLAS and CMS pixel detector upgrades for the HL-LHC~\cite{Collaboration:2285585,Collaboration:2272264}.  In particular, the sensors are $150$ $\mu$m thick with a pitch of $50\times 50$ $\mu$m$^2$.  The simulation of energy deposition, drift, and digitization is the same as in Ref.~\cite{Chen:2017yzr} and is briefly summarized here for completeness.  Charge deposition and straggling are provided by Geant4 using the \textsc{emstandard\_opt0} model\footnote{This is not accurate for thin sensors, but $200$ $\mu$m are sufficiently thick that the total deposited charge is well-modeled~\cite{Wang:2017ygj}.}.  The ionization energy is converted into electron-hole pairs assuming 3.6 eV/pair and electrons are transported to the collecting electrode, including drift and diffusion.  Collected electrons are digitized using a the time-over-threshold (ToT) method~\cite{603658}, with a linear charge-to-ToT conversion.  The analog threshold is varied, but the number of bits is fixed at 4, as suggested in Ref.~\cite{ben_optimal_digitize}.  Unless otherwise specified, the sensors are modeled without radiation damage.  The effects of radiation damage are approximated by reducing the collected charge according to the $n^+$-in-$n$ planar sensor results based on combining TCAD simulations from the Perugia~\cite{7935470} and New Delhi models~\cite{Dalal:2015rna} with drift, diffusion, and digitization presented in Ref.~\cite{Collaboration:2285585}.

%The analog threshold is 3000 electrons, there are 8 bits of ToT, and a minimum ionizing particle (MIP) at perpendicular incidence corresponds to a ToT of 128 (half of the available range). 

%-------------------------------------------------------------------------------
\section{Performance Metrics}
\label{sec:calculation}
%-------------------------------------------------------------------------------

Three important rates that are tied to the choice of threshold are the signal efficiency, the occupancy, and the noise rate.  The signal efficiency is the fraction of collected charge from a MIP.  Charge that diffuses to a neighboring pixel or is in the first or last pixel of a cluster may be below the threshold.  The threshold can also be used to control the overall hit rate in order to ensure that the occupancy is manageable.  For pixel detectors at the HL-LHC, the occupancy will be dominated by real hits and not noise.  However, the total occupancy still has a large contribution due to non-MIP hits.  Since it is difficult to accurately model the low-energy spectrum, instead of providing the total occupancy, we report the contribution of MIPs to the occupancy.  Finally, as the noise rate is well below the overall occupancy, it is important to report the error rate separately. 

\begin{figure}[h!]
\centering
\includegraphics[height=0.95\textwidth]{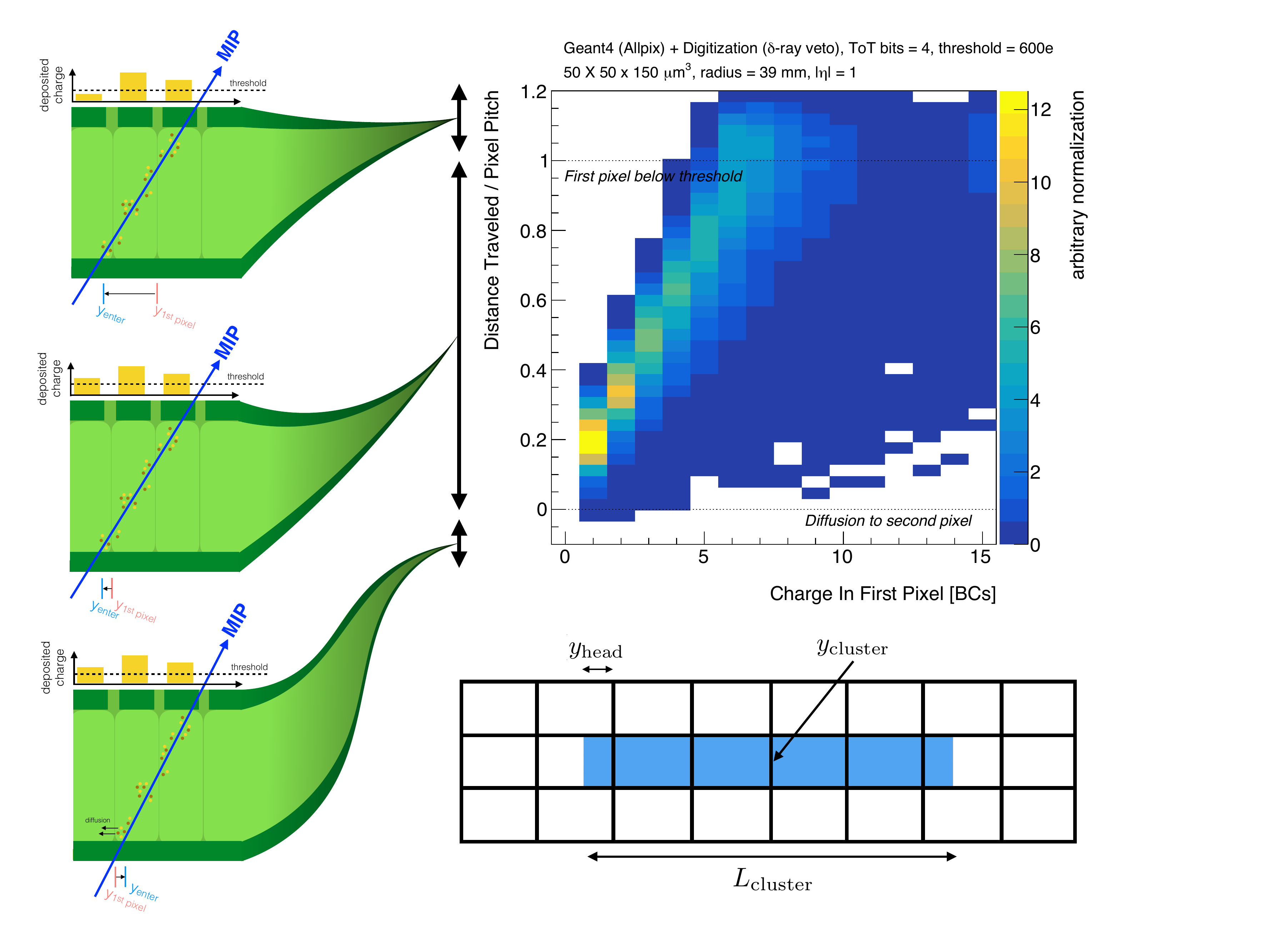}
\caption{A schematic diagram to illustrate the calculation of the position resolution metric.  The bottom right figure shows a pixel cluster, where the filled regions indicate the path of a MIP.  All of the information about the position and length of the cluster are contained in $y_\text{head}$ and $y_\text{tail}$; since the resolution of these two quantities should be approximately the same, we focus only on the former quantity.  The top right plot shows the distribution of the position traversed by a MIP normalized per bin of measured charge.  The left figures illustrate the definition of $\hat{y}$, which can be negative if enough charges diffusion into the previous pixel and can be more than $1$ if the first traversed pixel is below threshold.}
\label{fig:schematic}
\end{figure}

One of the important consequences of a reduced charge collection efficiency with increased threshold is that the estimated position resolution degrades.  Alongside the counting metrics described above, we also report the position resolution as a function of the threshold setting.   For a cluster of length $L_\text{cluster}$ like the one shown in the bottom of Fig.~\ref{fig:schematic}, all of the information about the position in the $y$ (long) direction as well as the longitudinal incidence angle is contained in $y_\text{head}$ - the location of the particle as it traversed the first pixel in the cluster.  As the tail and head position resolutions are approximately the same, the resolution on the position estimator $y_\text{cluster}=\frac{1}{2}(y_\text{head}+y_\text{tail})$ is $\sigma_{y_\text{head}}/\sqrt{2}$, while the resolution on the cluster length $y_\text{head}-y_\text{tail}$ is $ \sqrt{2}\sigma_{y_\text{head}}$.  Since the deposited charge scales with path length, one can use the amount of deposited charge in the first pixel to estimate the location $y_\text{head}$.   

The estimator for $y_\text{head}$ that minimizes the mean squared error is $\hat{y}_\text{head}(Q)=\langle y|Q\rangle$, where $Q$ is the (digitized) charge deposited in the first pixel.  The top right plot in Fig.~\ref{fig:schematic} shows the distribution of the distance traveled inside the first pixel against the first pixel's charge, at a threshold of 600e.  As expected, there is a linear increase in the distance traveled with increasing charge, until the MIP has passed through the entire sensor.  Due to significant straggling, there is a large spread in distance traveled for a given charge.  The average $\langle y|Q\rangle$ is computed assigning zero to the start of the pixel and normalizing by the pitch.  In rare circumstances, enough charge can diffuse to the pixel before the first traversed one in the pixel matrix and therefore, $\hat{y}$ can be negative.  In addition, the threshold can be sufficiently high that the first traversed pixel is below threshold and so $\hat{y}$ can also be bigger than $1$.  Each of these cases are illustrated schematically in the left diagrams of Fig.~\ref{fig:schematic}.  The position resolution is given by $\sqrt{\langle(\hat{y}_\text{head}-\langle\hat{y}_\text{head}\rangle)^2\rangle}$ and is approximately~\cite{binaryreadout} bounded by $\text{pitch}/\sqrt{12}$.

Pixels due to $\delta$-rays are excluded from the analysis because they register an anomalously high charge that has little to do with the position of the original MIP.   Especially for $\delta$-rays that travel many pixels before reaching their Bragg peak, the non-MIP signature can be identified and removed before estimating the MIP position.  The occurrence of $\delta$-rays for the first pixel in a cluster is about 1\%.

%For a cluster of hit pixels, the position estimator $\hat{y}$ that minimizes the mean squared error is given by $\langle 

%For each incidence angle we observe several distinct cluster shapes and for each shape we record the distribution of particle track entrance points to the sensor.  The optimal position estimator of each module given the shape and incidence angle is the mean of the entrance point distribution (the estimator that minimizes the mean squared error relative to the true position). Symbolically, the optimal position resolution is described by

%\noindent where $d\text{x}$ is defined in Fig.~\ref{fig:residuals}, $\Pr(s)$ is the probability of a particular shape s, $d\text{x}_s$ is conditioned on this shape $s$, and $\langle x_\text{true}|s\rangle$ is the average true entrance position given the shape $s$.  An analogous formula describes the optimal position resolution in the $y$ direction.  As a consequence of Eq.~\ref{eq:optimalres} , all that is needed to compute the optimal resolution is the RMS of the true position for all possible shapes.   Since our pixel information is binary, two clusters with the same hit pixel pattern are indistinguishable. We histogram the true entrance point independently for each cluster shape and compute the mean and RMS of the distribution. The mean is used as the optimal estimator for the position. As an example, Fig.~\ref{fig:histexample}\,(a) shows the true entrance point y coordinate histogram for the cluster shape shown in (b).  All position resolution results are shown normalized to the pitch/$\sqrt{12}$ assumption. 

%Occupancy and resolution.

\clearpage

%-------------------------------------------------------------------------------
\section{Threshold Schemes}
\label{sec:schemes}
%-------------------------------------------------------------------------------

We consider three schemes for setting charge thresholds:

\begin{itemize}
\item \textbf{Nominal}: If the charge is below the threshold, then the ToT is zero.  This is the usual way a fixed threshold is implemented: a comparator takes the output of the pixel pre-amplifier and compares it with a fixed threshold.
\item \textbf{$f_\text{share} = X\%$}: Pixel modules already exhibit a form of dynamic thresholds due to charge sharing (often called `cross-talk') via interpixel capacitance.  When a charge $q$ is deposited in one pixel, the neighboring pixels register $f_\text{share}q$, where $f_\text{share}$ depends on the capacitive coupling between pixels which in part scales with the length of the shared edge.  This means that the effective threshold for the neighbor of a hit pixel is reduced by $f_\text{share}q$~\cite{pixelbook}.  The value of $f_\text{share}$ is typically specified to be as small as possible, often being on the order of a few percent.  We propose to engineer $f_\text{share}$ to optimize the occupancy and resolution.  In practice, designing a pixel with a given $f_\text{share}$ while also simultaneously meeting other specifications may prove difficult, however, our goal is to study the impact of a larger $f_\text{share}$ so as to motivate future studies in a real chip.  Since we are assuming square pixels, we add $f_\text{share}q$ to the four neighbors sharing an edge and subtract $4f_\text{share}q$ from the primary pixel.  For the other schemes, the cross-talk is set to zero.
\item \textbf{$f_\text{neighbor} = X\%$}: Cross-talk is an indirect method for dynamic thresholds; instead, we propose to directly set the the threshold of a given pixel based on the activity in neighboring pixels.  The simplest such scheme is to have two thresholds: a nominal high threshold and a lower threshold that is $f_\text{neighbor}$ of the high one.  If any pixel is above the high threshold, then all of its neighbors see a lower threshold.  In practice, this would require explicit information sharing between pixels and may require significant added capacitance and/or power.  However, Sec.~\ref{sec:results} will show that this is a powerful scheme for maintaining both high efficiency and good position resolution. 
\end{itemize}

Before presenting results, we note that the latter two schemes affect only a fractionally small number of pixels for any given event, and therefore have a negligible impact on the overall noise rate.

\section{Results}
\label{sec:results}

\begin{figure}[h!]
\centering
\includegraphics[height=0.49\textwidth, valign=t]{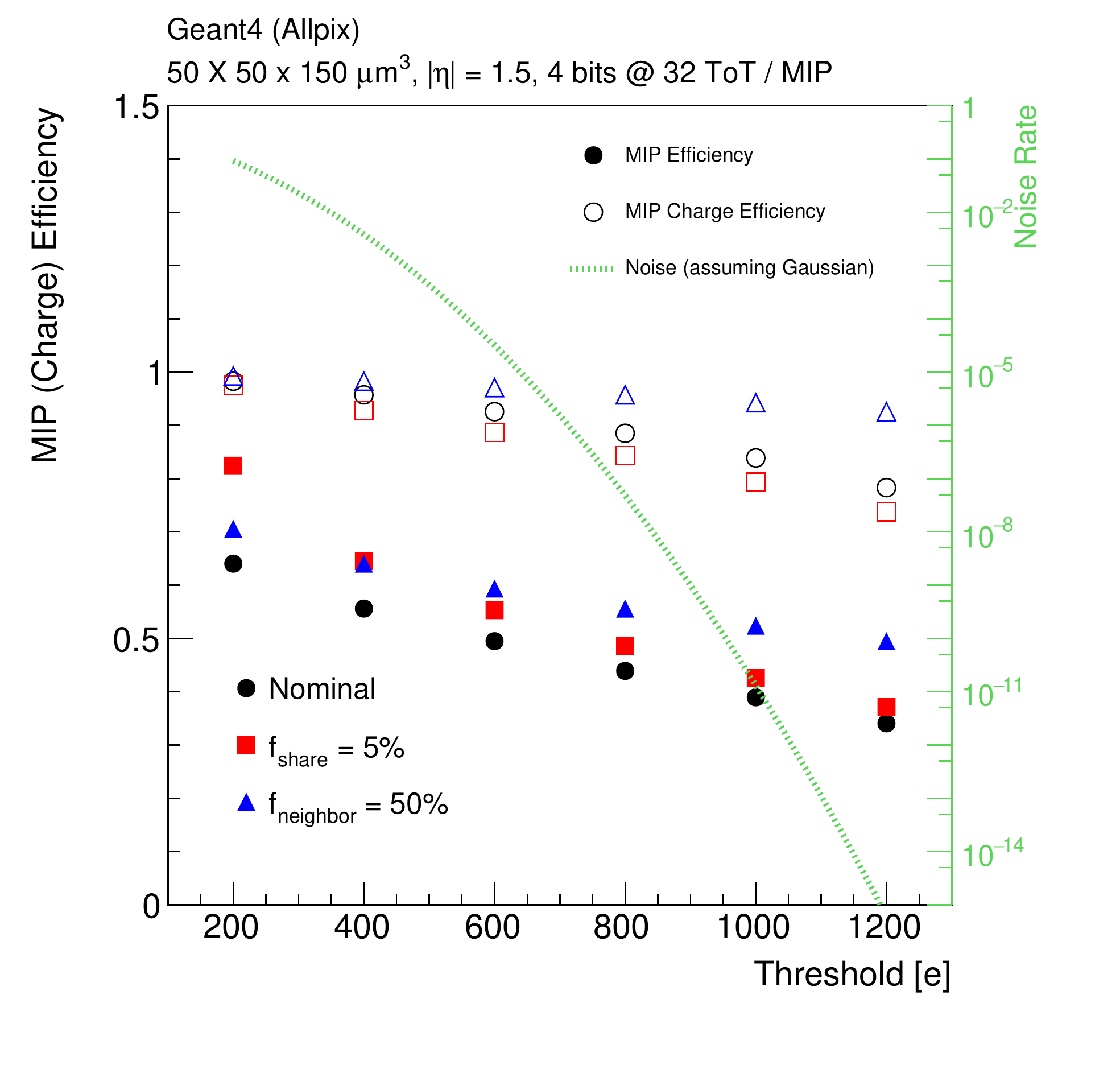}
\includegraphics[height=0.45\textwidth, valign=t]{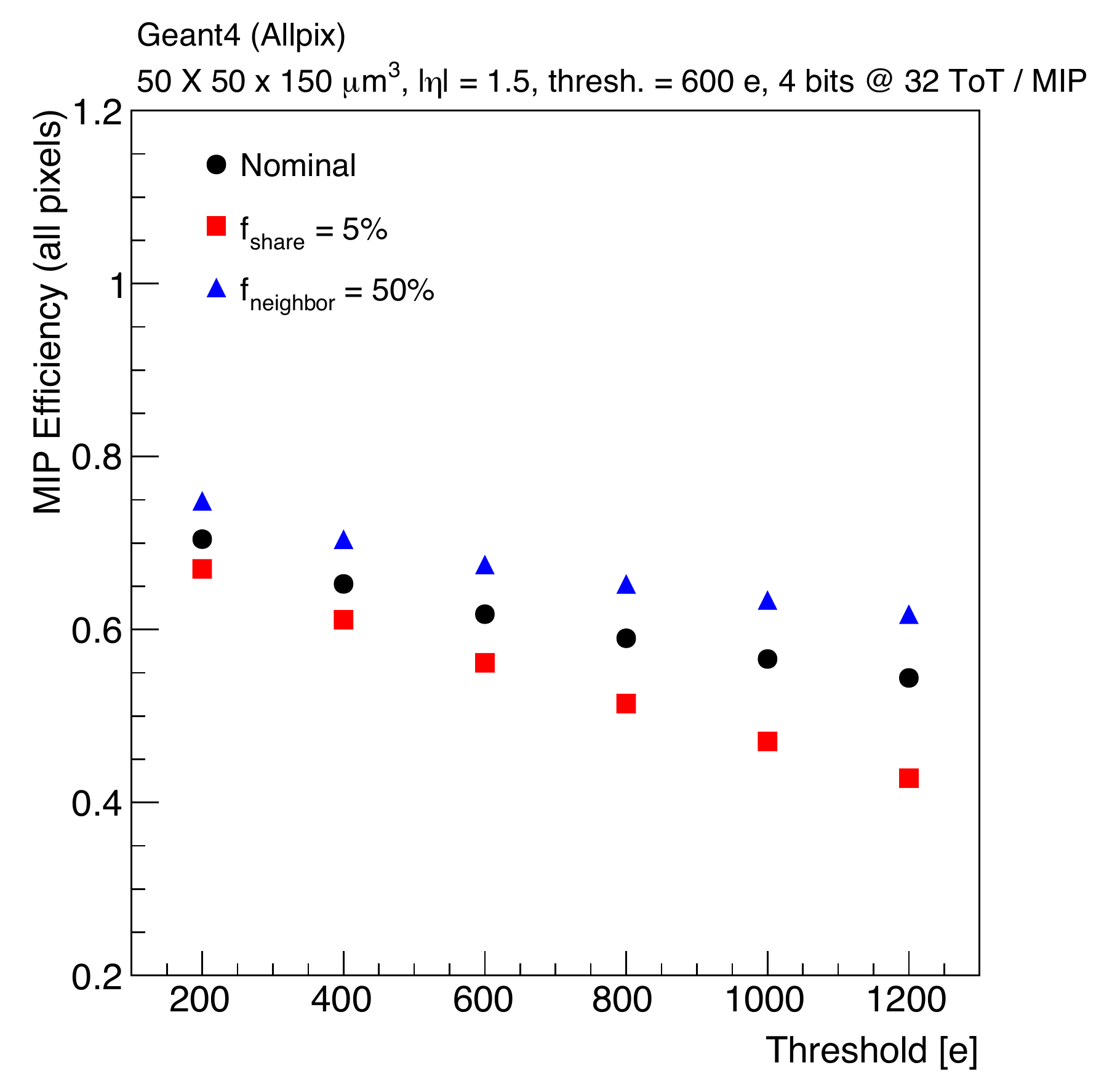}
\caption{Left: The MIP (charge) efficiency as a function of the threshold for the three threshold schemes; for reference, the noise rate is also given assuming an ideal 150e Gaussian noise profile.  Right: The MIP Effeciency measured over all pixels in which charge was deposited or shared.  In both cases, MIPs are incident a slight angle (corresponding to $\eta=1.5$) in order to increase the pixel multiplicity in clusters along the longitudinal direction.  The ToT is tuned so that a MIP at perpendicular incidence would correspond to a ToT of 32 if 15 were not the maximum value.}
\label{fig:resvthresh}
\end{figure}

Figure~\ref{fig:resvthresh} shows the MIP efficiency and charge efficiency for first traversed pixel in a cluster, as well as the MIP effeciency measured over all pixels,  as a function of the threshold for the three schemes introduced in Sec.~\ref{sec:schemes}.  In the case where the efficiencies are only given for the first traversed pixel in the cluster, the MIP charge efficiency is much higher than the efficiency to register any hit; this is a consequence of the fact that the charge in the first pixel is small when the path length is short.  As expected, increasing the threshold degrades both the (charge) efficiency.  For the chosen values of $f_\text{share}=5\%$ and $f_\text{neighbor}=50\%$, the hit efficiency is improved for every threshold.  The $f_\text{neighbor}$ scheme also has a higher MIP charge efficiency than the nominal approach. 

Additionally, the $f_\text{share}$ approach appears to have a lower MIP charge efficiency than the nominal approach, but this is an artifact caused by the increased charge from the neighbor as after digitization, it cannot be distinguished from the primary charge.   Notably,  the MIP efficiency is 5-10\% higher with the new threshold schemes.  The plot on the right of Fig.~\ref{fig:resvthresh} essentially shows the average fractional amount of pixels which go over threshold in a given scheme. As expected, this shows the same trend, but with the cross-talk scheme causing an increased rate of hit-losses relative to the other schemes (an effect which manifests most significantly on the edges of clusters, and is thus supressed in the plot on the left).

For reference, the left plot of Fig.~\ref{fig:resvthresh} also shows the noise rate, assuming ideal Gaussian noise, in which the rate decreases exponentially with increasing threshold.  In practice, the noise is not exactly Gaussian, and the suppression with increased threshold is not as strong as indicated.  However, the fact that the noise rate is still significantly suppressed with increasing threshold, coupled with the trends shown in Fig.~\ref{fig:resvthresh}, indicate that it is possible to have a higher threshold without compromising the MIP (charge) efficiency.

\begin{figure}[h!]
\centering
\includegraphics[width=0.49\textwidth, valign=t]{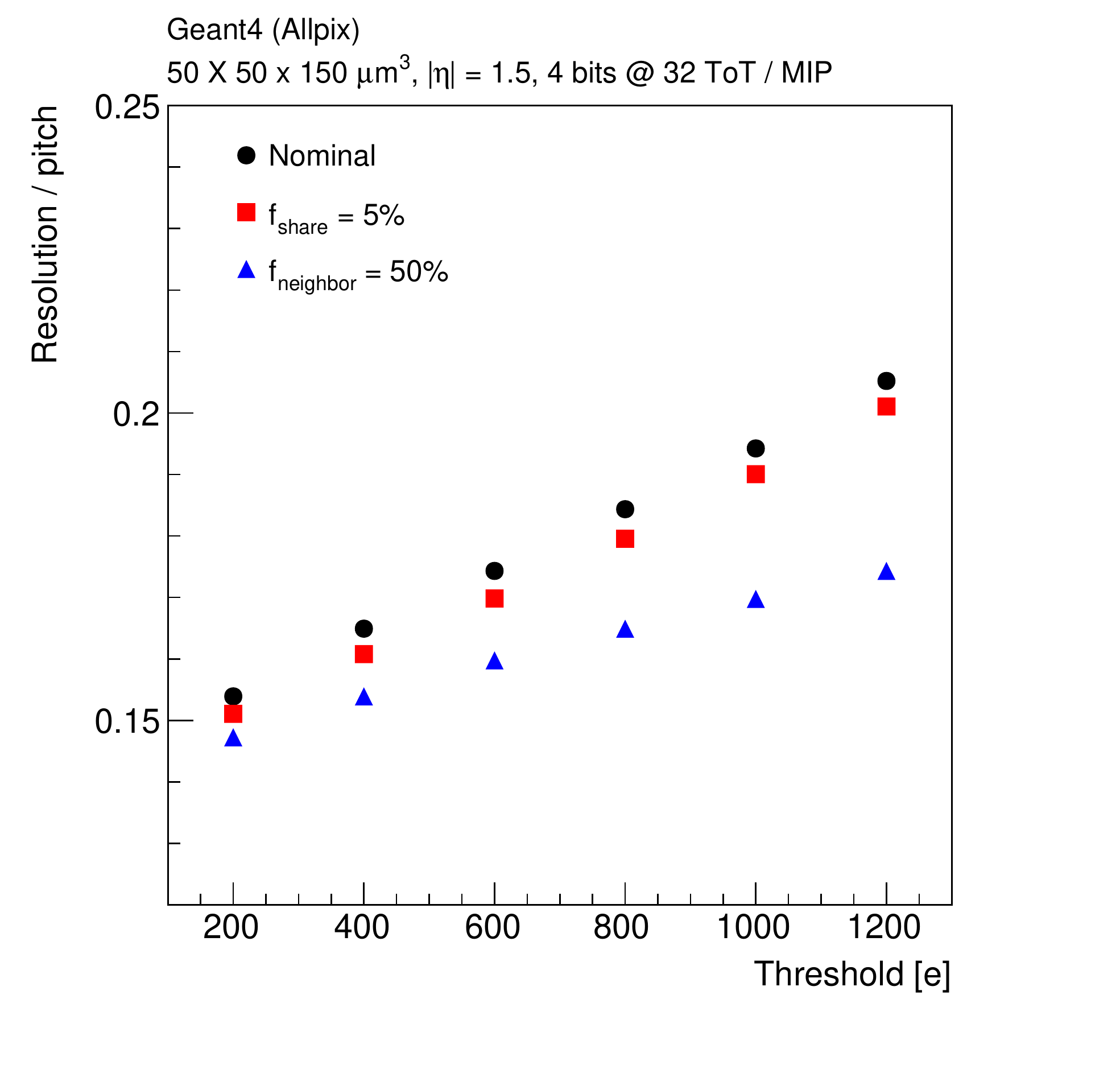}
\includegraphics[width=0.49\textwidth, valign=t]{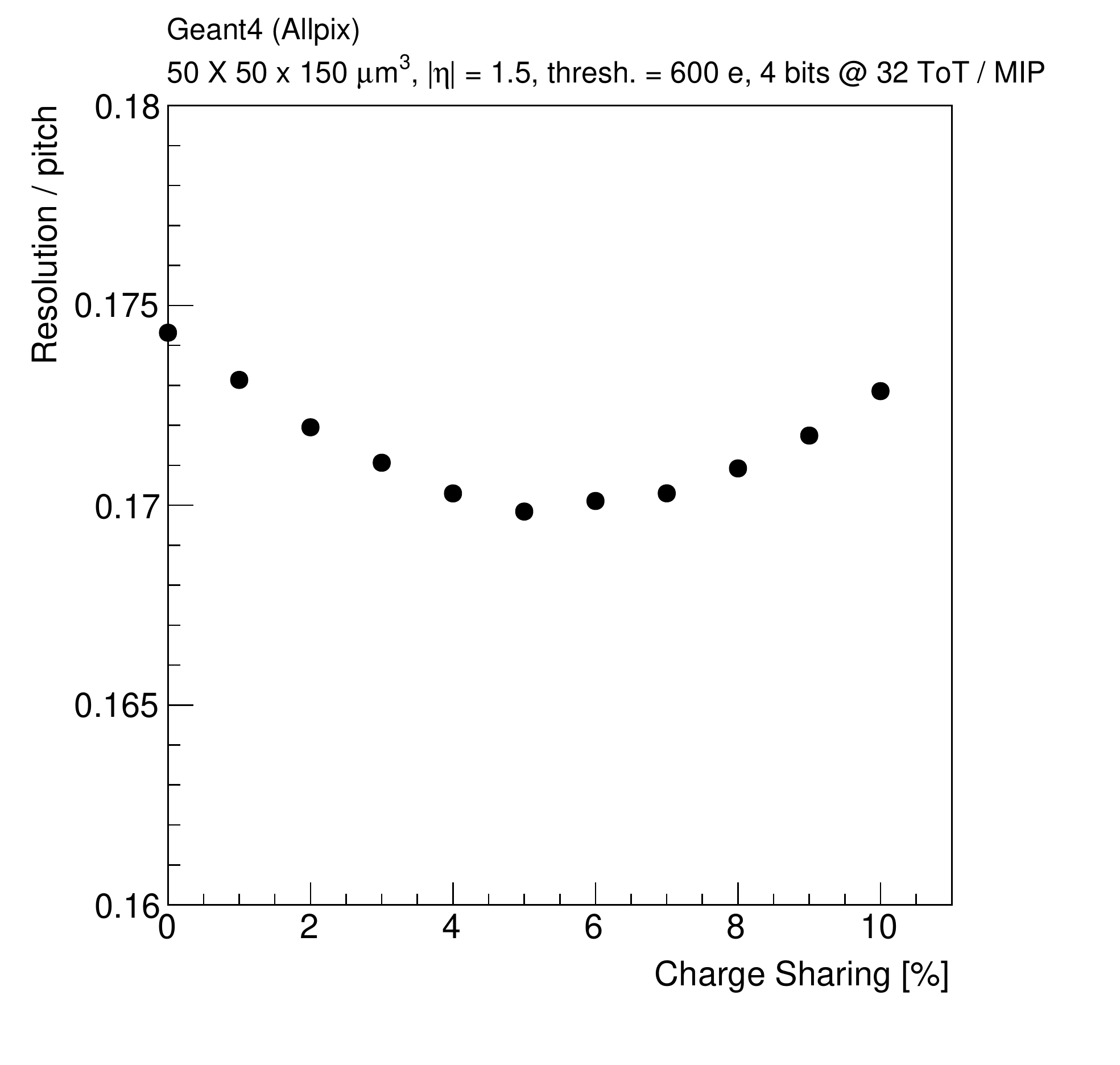}
\caption{Left: the $y_\text{head}$ position resolution as a function of the threshold for the three schemes.  Right: The $y_\text{head}$ position resolution as a function of the charge sharing ($f_\text{share}$). In both cases MIPs are incident at $\eta=1.5$, with the same charge tuning as in Fig.~\ref{fig:resvthresh}.}
\label{fig:scanshare}
\end{figure}

Fig.~\ref{fig:scanshare} contains two plots: the first illustrating the variation of position resolution with threshold in the three schemes under investigation, and the second showing the resolution as a function of $f_\text{share}$. 

Focussing first on the left plot, we see that the two new schemes improve the resolution for all values of threshold, and that the resolution worsens with increasing threshold, akin to the MIP efficiency. Furthermore, with a value of $f_\text{neighbor}=50\%$, the triangle points in the left plot are the same as the nominal points with a threshold reduced by 50\%.  The improvement from the $f_\text{share}=5\%$ scheme is more modest, but is still a few percent for all thresholds.  Additionally, the shallow trend of the direct-talk scheme implies that increased thresholds could be applied with relatively less detriment to the resolution than in the other two schemes..

The right plot  highlights the sensitivity of the resolution to the exact amount of cross-talk. Interestingly, there is an optimal amount of charge sharing at 5\% for the given incidence angle, pitch, threshold, and charge tuning.  This is to be expected, as increasing $f_\text{share}$ from zero improves the resolution until information about the charge from the first pixel is washed out by the contribution from the neighbor that went over the threshold.  The absolute change in the resolution is about 2\%, but subtracting in quadrature, the additional resolution is about 20\%. 

\begin{figure}[h!]
\centering
\includegraphics[width=0.5\textwidth]{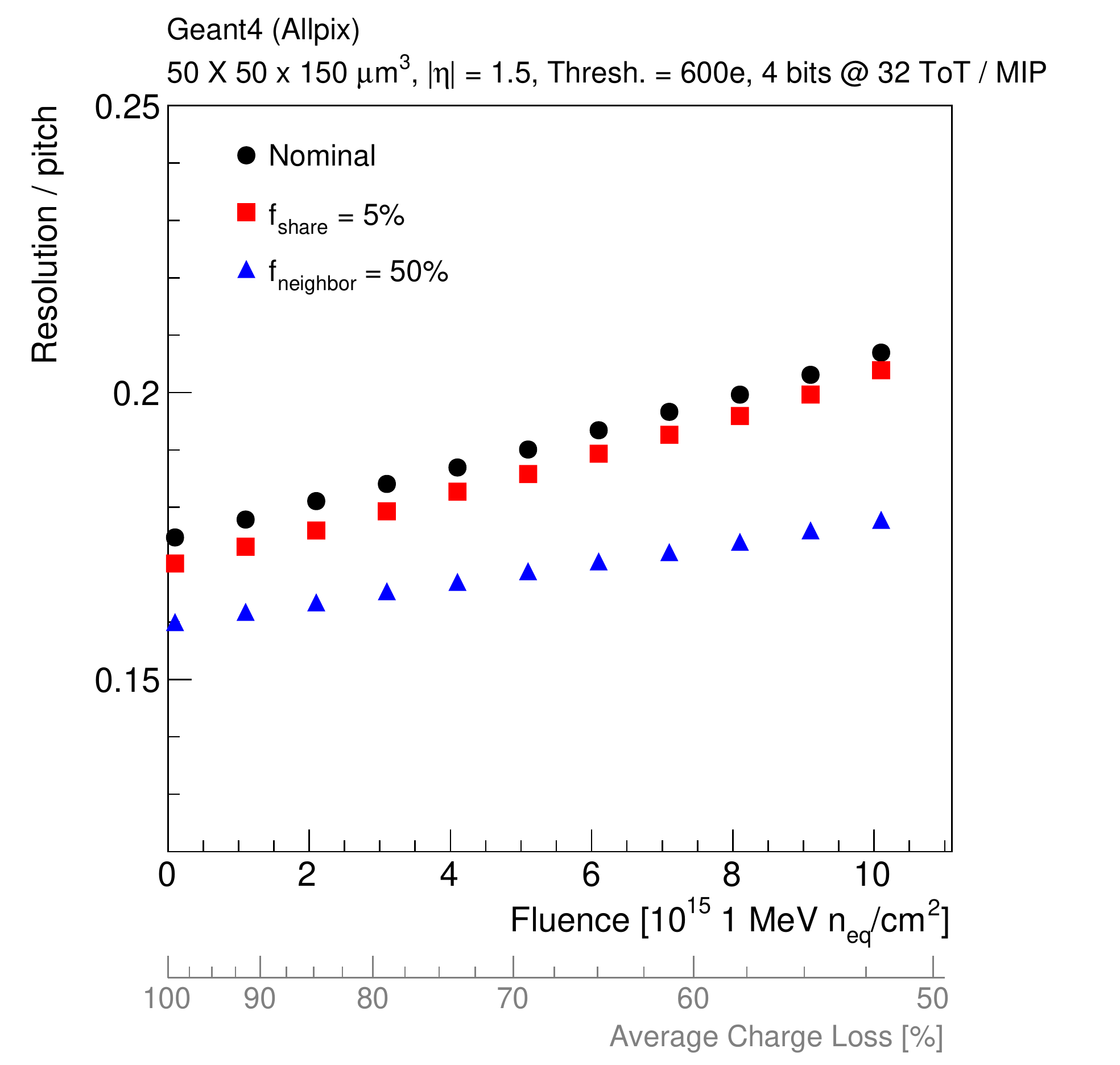}
\caption{The position resolution as a function of the silicon 1 MeV $n_\text{eq}/\text{cm}^{2}$ fluence for the three threshold schemes at a fixed threshold of 600e and with the same tuning as Fig.~\ref{fig:resvthresh}.  The average charge loss from Ref.~\cite{Collaboration:2285585} is given as a second axis.}
\label{fig:raddamage}
\end{figure}

The intense radiation environment of current and future hadron colliders is one of the greatest challenges for silicon-based pixel detectors.   Figure~\ref{fig:raddamage} shows the position resolution as a function of the non-ionizing energy loss for a fixed threshold.  Since charge is lost from charge trapping, the resolution degrades with fluence.  The innermost layers of the HL-LHC detectors will need to cope with about $10^{16}$ 1 MeV $n_\text{eq}/\text{cm}^{2}$.  Given the assumptions going into Fig.~\ref{fig:raddamage}, the $f_\text{neighbor}=50\%$ scheme has the same position resolution after the full HL-LHC fluence as the nominal scheme does with an unirradiated sensor. The clear superiority of the direct scheme with respect to radiation hardness is unsurprising as it is the least severely affected by changes in signal-size which are not significant enough to drive pixels under threshold.

The results presented thus far were based on the simulation of pixels with symmetric side lengths (i.e. square).  However, there has been considerable investigation of the potential use of asymmetric pixels at the LHC, such as $25\times 100$ $\mu$m$^2$, which can trade off the position resolution in the longitudinal direction ($z_0$) for increased resolution in the transverse direction ($d_0$), which is more important for flavor tagging. Importantly, when pixels are not square, the charge sharing will not be the same for the long and the short sides.  While the complete calculation of charge sharing is complicated and sensor-specific, the capacitance (and thus the charge sharing) is approximately proportional to the side length of the pixel.  For example, in the $25\times 100$ $\mu$m$^2$ case, the short sides will exhibit $4$ times less charge sharing than the long sides.  

\begin{figure}[h!]
\centering
\includegraphics[width=0.45\textwidth]{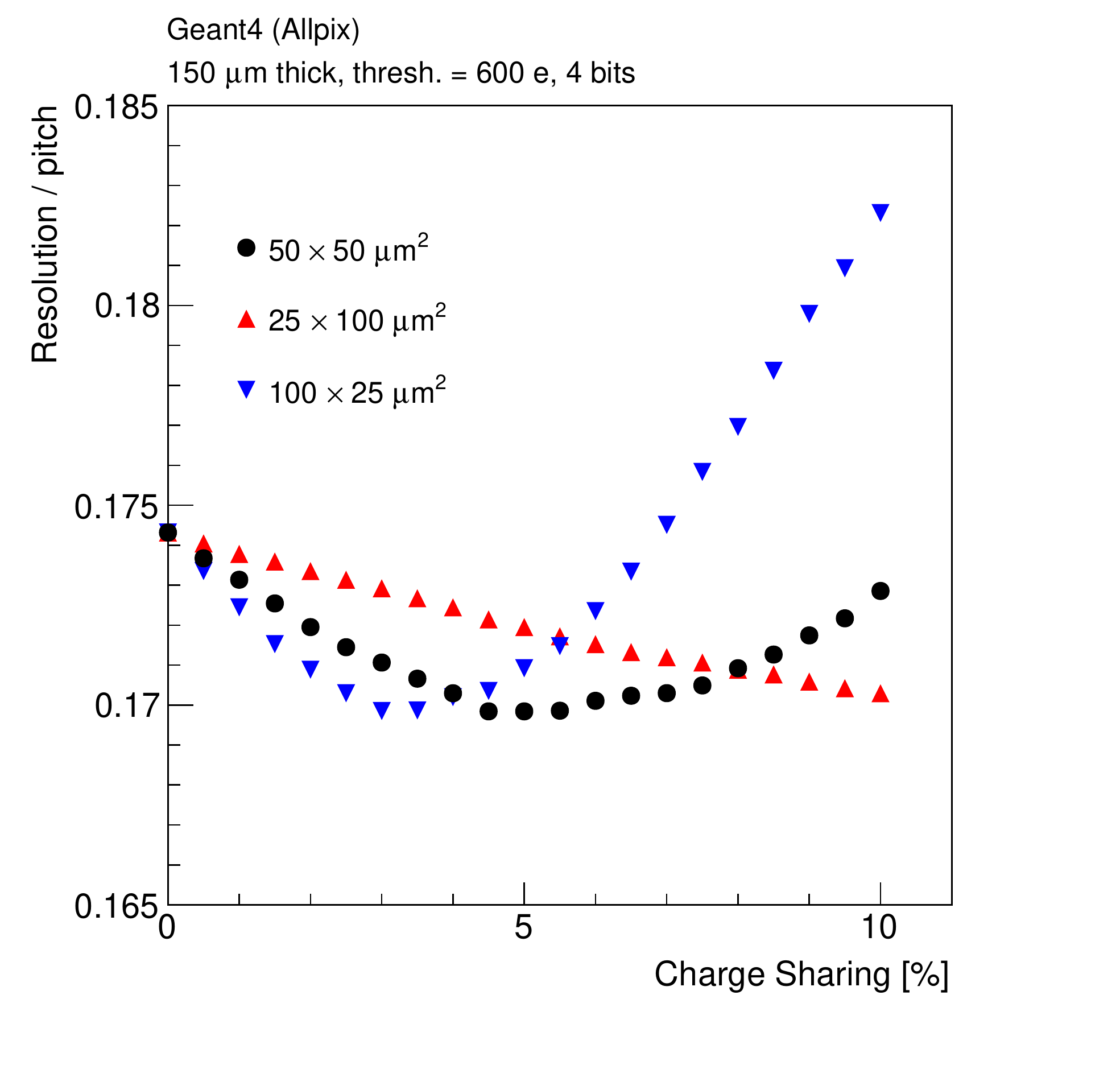}\includegraphics[width=0.45\textwidth]{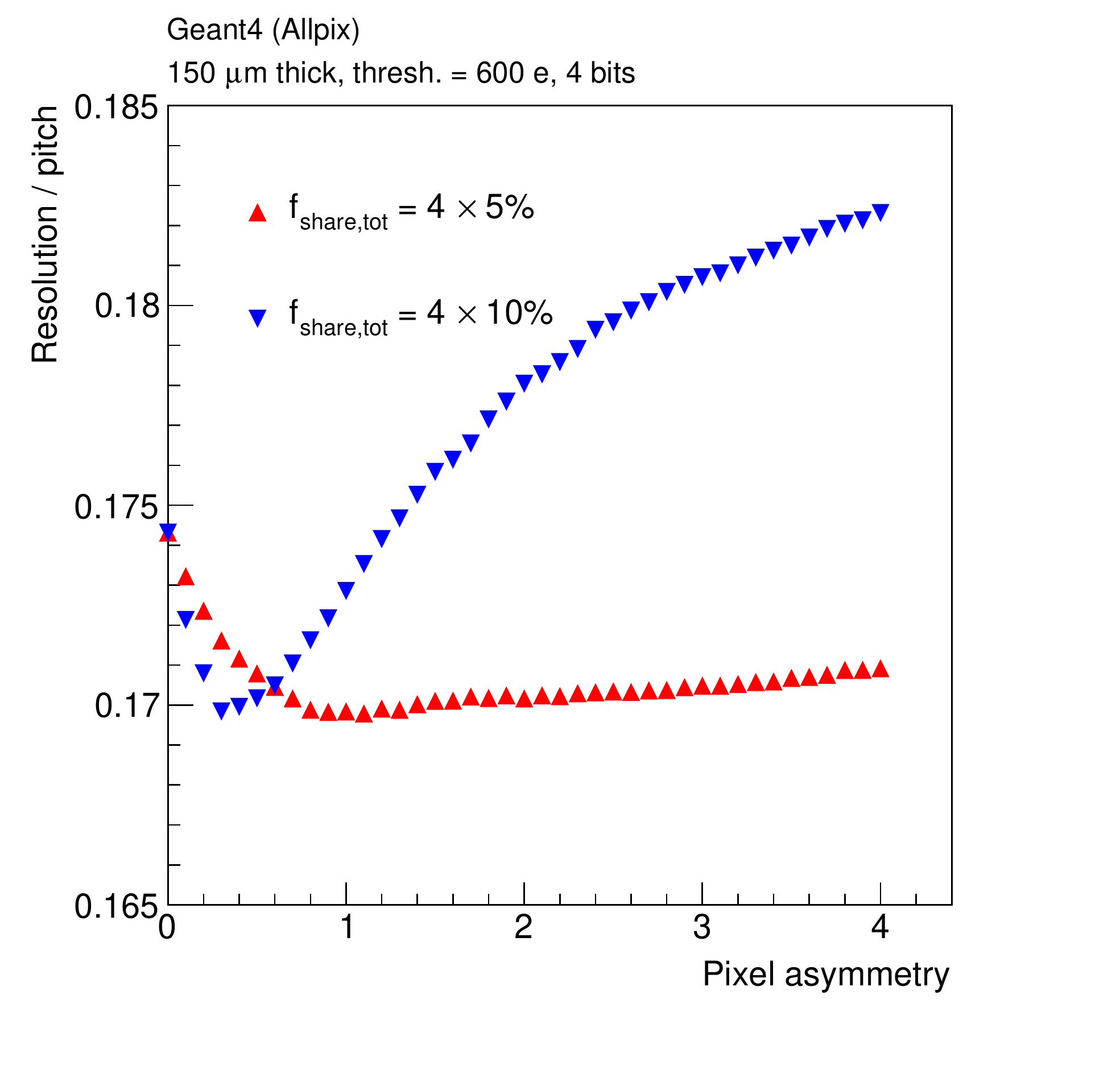}
\caption{Left: the relative position resolution as a function of the amount of charge sharing for three different pixel configurations (the resolution always corresponds to the first dimension given in the legend).  Right: for a fixed amount of total charge sharing (20\% charge loss from the primary pixel), the position resolution is shown as a function of the asymmetry in the pixel pitches.  A value of 0.25 corresponds to $25\times 100$ $\mu$m$^2$.}
\label{fig:asymmetricpixels}
\end{figure}

Figure~\ref{fig:asymmetricpixels} illustrates how the position resolution changes with asymmetric pixels.  In order to control for effects related to the actual amount of charge deposited due to the pixel size, all results are actually simulated with the $50\times 50$ $\mu$m$^2$ setup from earlier.  However, the amount of sharing in the $x$ and $y$ directions is now different and is set proportional to the side length.  If the sharing before was $f_\text{share}$, then the new sharing is $f_\text{share}'=2f_\text{share}/(1+\text{pixel asym.})$, which is chosen so that the total charge lost by the primary pixel is still $4f_\text{share}$.  The pixel asymmetry is the ratio of the transverse to longitudinal pixel pitch.  

The left plot of Fig.~\ref{fig:asymmetricpixels} shows how the relative resolution changes for different configurations as a function of the amount of charge sharing.  A charge sharing value of 5\% means that the primary pixel loses $4\times 5\%$ of its charge to its neighbors, divided up in a way that is proportional to the shared side length.  When the pitch is smaller, the optimal amount of charge sharing increases.  In the $25\times 100$ $\mu$m$^2$ configuration, the optimal sharing for the long side is about 3\% while there is no optimal value for the short side (larger value than the 10\% cutoff is desired).   The right plot of Fig.~\ref{fig:asymmetricpixels} fixes the total charge sharing and varies the pixel asymmetry.  For a total charge sharing of 5\%, the down-triangles and circles from the left plot of Fig.~\ref{fig:asymmetricpixels} are nearly the same, which is consistent with the broad minimum in the right plot for the up-triangles.  In contrast, there is a strong dependence on the pixel asymmetry in the sub-optimal case of 10\% charge sharing.  

%This behaviour is not unsurprising, as a reduction in the total amount of charge sharing correspondingly leads to a decrease in the affect any asymmetry variations might have on the overall resolution. Why the low antisymmetry behviour though - give nthat we're controlling for deposited charge- presumably just because close to 0 we're essentially turning off charge sharing in  one of the dimensions - and any amount of charge sharing is better than none?

\section{Discussion}
\label{sec:disc}

Using capacitive coupling to implement the dynamic threshold has the advantage that the information from the primary hit is transferred nearly instantly to the neighboring pixels.  The disadvantage is that designing a specific amount of capacitive coupling is challenging, especially given the tight constraints from other design requirements (including noise and power consumption).

\begin{figure}[h!]
\centering
\includegraphics[width=0.6\textwidth]{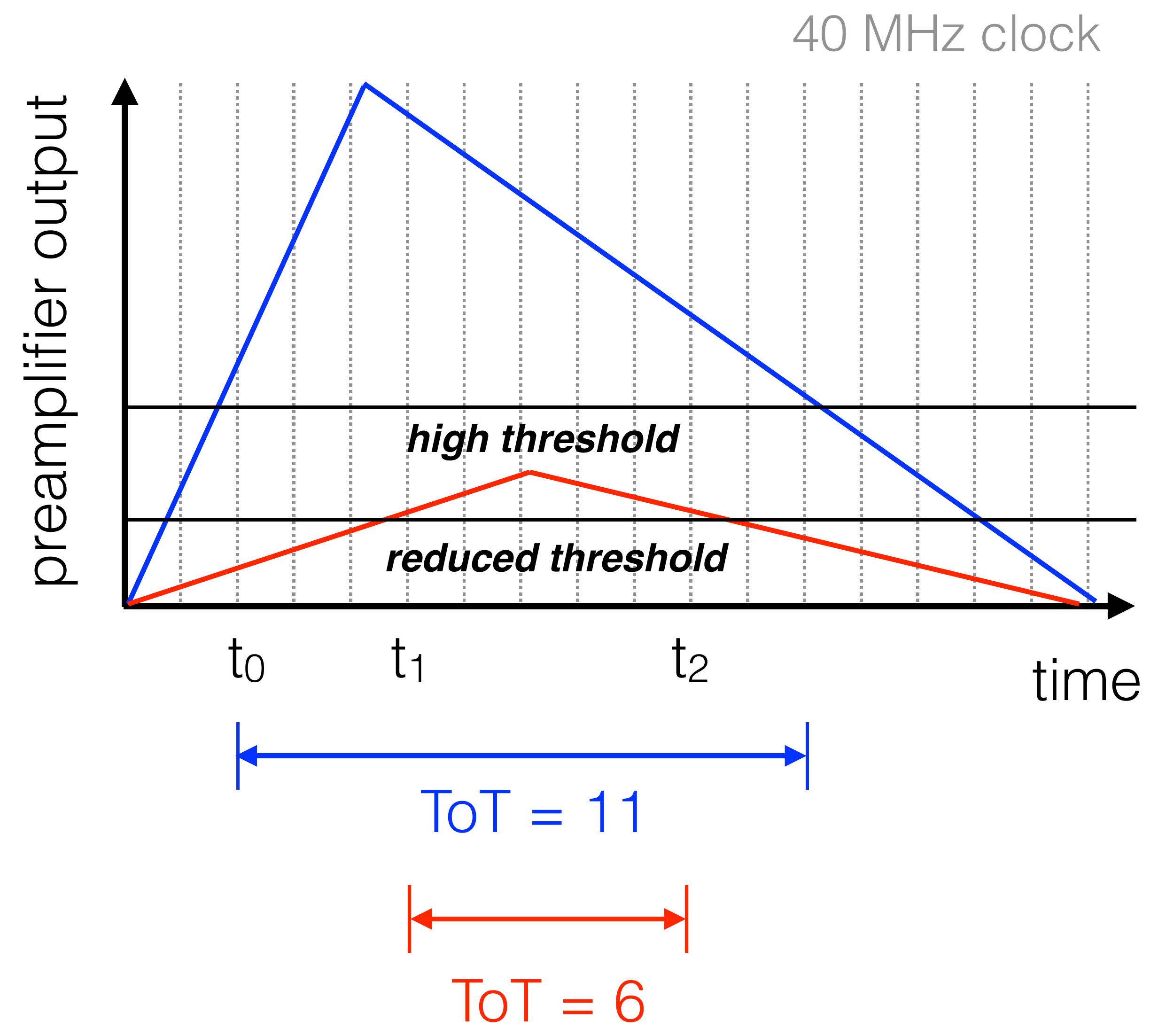}
\caption{A schematic illustration of the charge digitization for two particles going through neighboring pixels in the same bunch crossing: one with a large charge (blue) and one with a small charge (red).  The small charge particle does not pass the initial high threshold, but would pass the reduced threshold if the new threshold could be set before $t_2$.  The ToT for the small charge in this scheme would be at most 6, when the threshold is reduced by $t_1$. }
\label{fig:timing}
\end{figure}

In the alternative scheme where active logic is used to reduce the threshold on the neighbors, information must be quickly sent to the neighboring pixels.  Figure~\ref{fig:timing} illustrates this time constraint when one hit passes the initial high threshold and a neighboring hit would only pass a reduced threshold.  The first clock cycle where this hit is recorded to be above the high threshold is $t_0$ and the first clock cycle for which the smaller hit would be above the reduced threshold is $t_1$, while it goes below this threshold at $t_2$.  This second hit will be recorded as long as the second threshold can be reduced in a time $t_2-t_0$.  The charge resolution of the second hit will be optimal when the communication time is only $t_1-t_0$; any will result in a degraded resolution.

\clearpage

%-------------------------------------------------------------------------------
\section{Conclusions}
\label{sec:concl}
%-------------------------------------------------------------------------------

The HL-LHC presents significant challenges for pixel module design and now is the time to consider new possibilities for optimizing the information saved for offline analysis.  We have presented two schemes for dynamic thresholds which use information from neighboring pixels in order to increase the MIP efficiency, with little or no increase in the noise rate.  One scheme exploits the natural interpixel capacitance cross-talk to lower the effective threshold of pixels neighboring those with a large charge deposition.  Furthermore, from the perspective of position resolution, there is an optimal amount of charge sharing.  While in practice it may be difficult to engineer a particular level of charge sharing, given other specifications, these results suggest that design studies are worthwhile, especially in light of the challenges posed by the LH-LHC conditions.  One drawback of the capacitive coupling scheme is that the effective decrease in the threshold is random and varies significantly with the straggling of MIP charge depositions. 

Secondly, we propose a scheme which instead utilizes two fixed thresholds, thus circumventing the aforementioned challenge.  This algorithmically simple scheme presents a lower threshold to all pixels next to a pixel that passes a high, nominal threshold.  This procedure significantly improves the resolution and MIP efficiency, but practical implementations would depend on a mechanism for rapid communication between neighboring pixels.  Indeed, adding circuitry for this purpose would certainly increase the capacitance and/or the power consumption, so such tradeoffs require a thorough investigation.  

Signal efficiency and position resolution are crucial for both track reconstruction and flavor tagging at the LHC, and thus it is conceivable that the trade-offs of the proposed dyamic threshold schemes may be outweighed by the gains.  Certainly, considerable amounts of potentially useful information are present in the neighborhood around pixels which are not being explicitly used, and which could significantly improve detector performance for the HL-LHC and beyond. 

\section{Acknowledgments}

We would like to thank Maurice Garcia-Sciveres and Timon Heim for many useful discussions as well as the RD53 collaboration for encouragement and feedback.  This work was supported by the U.S.~Department of Energy, Office of Science under contract DE-AC02-05CH11231.

\bibliographystyle{elsarticle-num}
\bibliography{myrefs.bib}{}

\end{document}